\input{aipcheck}

\documentclass[
    ,final            
  ]
  {aipproc}

\layoutstyle{6x9}
\usepackage{amsmath,amssymb, amsthm}
\usepackage{graphicx,epsf}
\def\half{\frac{1}{2}}

\def\rf#1{(\ref{#1})}
\newcommand{\eqa}{\begin{eqnarray}}
\newcommand{\neqa}{\end{eqnarray}}
\newcommand{\bea}{\begin{eqnarray}}
\newcommand{\eea}{\end{eqnarray}}
\newcommand{\be}{\begin{equation}}
\newcommand{\ee}{\end{equation}}

\newcommand{\Ref}[1]{(\ref{#1})}

\renewcommand{\texttt}{{}}
\usepackage{graphicx}

\def\sdmass{\ensuremath{\sqrt{a_0}/2}}
\def\su2{\ensuremath{\mathfrak{su(2)}}}
\def\SU2{\ensuremath{{SU(2)}}}
\begin{document}


\title{%
In, Through and Beyond the Planck Scale} 
\classification{
                }
\keywords      {}
\author{Leonardo Modesto}{ address={Perimeter Institute for Theoretical Physics, 31 Caroline St.N., Waterloo, ON N2L 2Y5, Canada}}

\author{Isabeau Pr\'emont-Schwarz}{ address={Perimeter Institute for Theoretical Physics, 31 Caroline St.N., Waterloo, ON N2L 2Y5, Canada}}


\begin{abstract} \noindent
In this paper we have recalled the semiclassical metric
obtained from a classical analysis of the loop quantum black hole
(LQBH). We show that the regular metric 
is self-dual: 
the form of the metric is invariant under the exchange $r \rightarrow a_0/r$
where $a_0$ is proportional to the minimum area in LQG. 
Of particular interest, the symmetry imposes that if an observer in $r\rightarrow +\infty$ sees a black hole of mass $m$
an observer in the other asymptotic infinity beyond the horizon (at $r \approx 0$) sees a dual mass
$m_P^2/m$. We then show that small LQBHs are stable and could be a component of dark matter.
Ultra-light LQBHs created shortly after the Big Bang
would now have a mass of approximately $10^{-5} \, m_P$
and emit radiation with a typical energy of about $10^{13} - 10^{14} {\rm eV}$ but they would also emit cosmic rays of much higher energies, albeit few of them. 
If these small LQBHs form a majority of the dark matter of the Milky Way's Halo,
the production rate of ultra-high-energy-cosmic-rays (UHECR) by these ultra light black holes would be compatible with the observed rate of the Auger detector.

\end{abstract}

\maketitle

%

\section*{Introduction}

Black holes 
are an interesting place for testing the validity of 
 ``Loop quantum gravity" (LQG) \cite{book}. 
In the past years, inspired by ``loop quantum cosmology", 
applications of LQG ideas to the Kantowski-Sachs space-time
lead to some interesting results. 
In particular, it has been
shown 
\cite{work2} 
that it is
possible to solve the black hole singularity problem by using
tools and ideas developed in the full LQG. 
There is also work of a semiclassical nature which tries to solve the black hole singularity
problem \cite{RNR}. 
 In these papers the author use an effective Hamiltonian constraint
 obtained by replacing the Ashtekar connection $A$ with the holonomy $h(A)$
 and they solve the classical Hamilton equations of motion exactly
 or numerically.
 The main result is that
 the minimum area 
 of full LQG is the
 fundamental ingredient to solve
 the black hole space-time singularity problem at $r=0$.
 The $S^2$ sphere bounces on the minimum area $8 \pi a_0$ 
 and the singularity disappears (
the Kretschmann  invariant is regular in all of space-time). 

This paper is organised as follows.  In the first section we
recall the semiclassical black hole solution obtained in \cite{RNR} 
and we show the self-duality property of the metric.
We take special notice of ultra-light black holes which differ qualitatively from 
Schwarzschild black holes even outside the horizon. We show that their horizons are hidden behind a wormhole of Planck diameter.
In the second section we study the phenomenology of LQBHs. 
We show ultra-light LQBHs can solve the dark matter problem and
simultaneously to be the missing source for the ultra-high-energy-cosmic-rays  (UHECRs).

\section{Regular and Self-Dual Black Holes} 
In this section we recall the 
semiclassical black hole solution obtained recently in 
$r \leqslant 2m$ \cite{RNR}, \cite{work2}. 
The semiclassical metric is
\begin{eqnarray}
 ds^2 = -\frac{ (r - r_+) (r-r_-)(r+ r_{\star})^2 }{r^4 + a_0^2}dt^2 
 +\frac{(r+ r_{\star})^2 (r^4 + a_0^2) \, dr^2}{(r-r_+)(r-r_-)r^4} 
 %
+\Big(\frac{a_0^2}{r^2} +
r^2\Big) d\Omega^{(2)},
\label{metricabella}
\end{eqnarray}
where
$r_+=2m$, $r_- = 2m {\mathcal P}(\delta_b)^2$, $r_{\star} = 2m {\mathcal P}(\delta_b)$,
$a_0 = A_{\rm Min}/8 \pi$ and $A_{\rm Min}$ is the minimum area of LQG. ${\mathcal P}(\delta_b)$
is a function of the polymeric parameter $\delta_b$ \cite{RNR},
${\mathcal P}(\delta_b) =[(1+\gamma^2 \delta_b^2)^{1/2} -1]/[(1+\gamma^2 \delta_b^2)^{1/2} +1]$.
The area operator in LQG has a discrete spectrum, irreducible units of area --- associated to an edge on a spin-network ---  in LQG have area $A(j):= 8\pi\gamma \sqrt{j(j+1)} l_P^2$ where $\gamma$ is the Immirzi parameter believed to be $\gamma=0.2375$ \cite{gamma}, $j$ is a half-integer labelling an irreducible representation of \SU2 and $l_P$ is the Planck length. Looking at this, it is natural to assume that the minimum area in LQG is $A_{\rm Min}=A(1/2)= 4\pi\gamma \sqrt{3} l_P^2\approx 5 l_P^2$. One should however not take this exact value too seriously for various reasons 
\cite{RNR} and 
parameterizes this ignorance with a parameter $\beta$ and define
$A_{\rm Min}= \beta A(1/2)= 4\pi\gamma\beta \sqrt{3} \, l_P^2\approx 5\beta l_P^2$, and so $a_0=A_{\rm Min}/(8\pi)= \gamma\beta \sqrt{3}\, l_P^2/2\approx 0.2\beta l_P^2$ where the expectation is that $\beta$ is not many orders of magnitude bigger or smaller than $1$, in this article we mostly consider $\beta\approx 1$ or $\beta=4$ when more precision is need, but in the end the precise choice of $\beta$ does not change much.

\begin{figure}
 %
  \includegraphics[height=6.0cm]{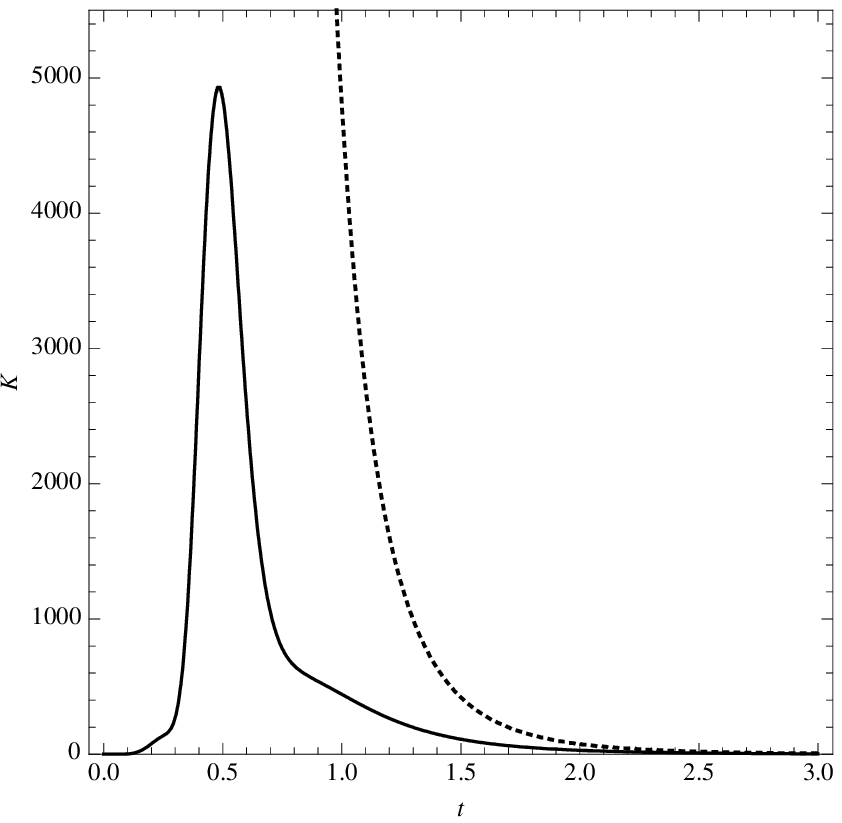} \hspace{1cm}
  \includegraphics[height=7cm]{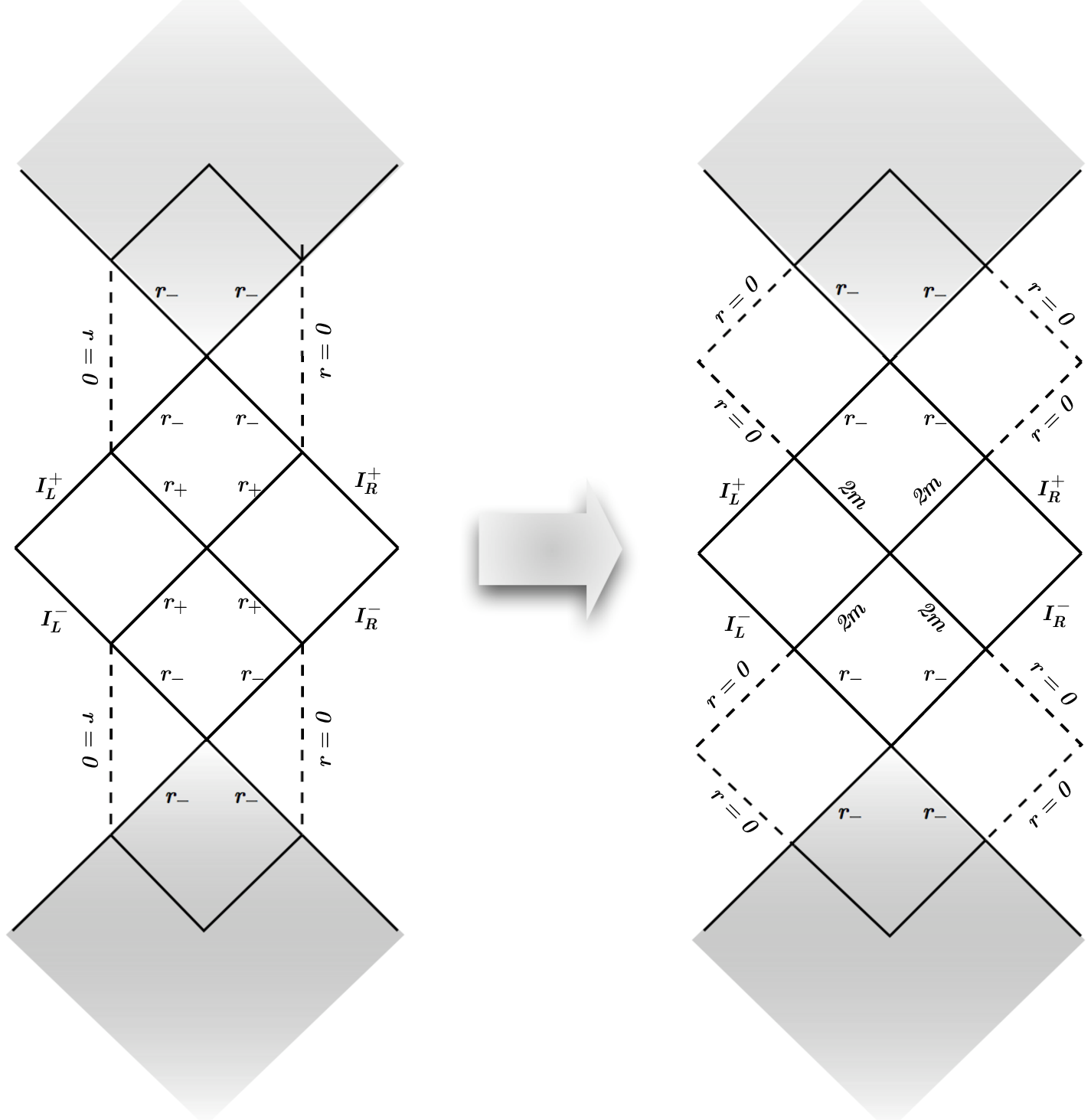}
  \caption{\label{penrose}
 %
 %
 %
 The picture on the left represents a plot of the
 Kretschmann  scalar
  invariant ${\rm R}_{\mu \nu \rho \sigma} {\rm R}^{\mu \nu \rho \sigma}$
for $m = 10$, $p_{b}^{0} =1/10$ and $\gamma \delta_b = \log(4)/\pi$, $\forall t \geqslant 0$; the large
 $t$ behaviour 
  is $1/t^6$.
  The picture on the right represents the maximal space-time extension of the LQBH on the right and the analog extension
  for the Reissner-Nordstr\"om black hole.}
  \end{figure}

The regular properties of the metric are: 
$\lim_{r \rightarrow +\infty} g_{\mu \nu}(r) = \eta_{\mu \nu}$, 
$\lim_{r \rightarrow 0} g_{\mu \nu}(r) = \eta_{\mu \nu}$,
$\lim_{m, a_0 \rightarrow 0} g_{\mu \nu}(r)  = \eta_{\mu \nu}$,
$K(g) < \infty \,\, \forall r$,
$r_{\rm Max}(K(g)) \propto l_P$.
Where $r_{\rm Max}(K(g))$ is the radial position where the Kretschmann invariant achieves its
maximum value. Fig.\ref{penrose} is a graph of $K$, it is regular in all of space-time and the large
$r$ behaviour is asymptotically identical to the classical singular scalar
${\rm R}_{\mu \nu \rho \sigma} {\rm R}^{\mu \nu \rho \sigma} = 48 m^2/r^6$.
The resolution of the regularity of $K$ is a non perturbative result, in fact for small values of the
 radial coordinate $r$,
$K\approx 3145728 \pi^4 r^6/A_{\rm Min}^4 \gamma^8 \delta_b^8 m^2$ diverges for
$A_{\rm Min}\rightarrow 0$
or $\delta_b \rightarrow 0$.
A crucial difference with the classical Schwarzschild solution is that
the 2-sphere $S^2$ has a minimum for
$r_{min} = \sqrt{a_0}$ 
and the minimum square radius is $p_{c}(r_{min}) = 2 a_0$.
The solution has a spacetime structure very similar to the
Reissner-Nordstr\"om metric because of the inner horizon in
$r_{-} = 2 m {\mathcal P}(\delta_b)^2$ (
 for $\delta_b \rightarrow 0$, $r_{-} \approx m \gamma^4 \delta_b^4/8$).
We observe that the position of the inside horizon is $r_-\neq 2m \,\, \forall \gamma \in\mathbb{R}$
(we recall that $\gamma$ is the Barbero-Immirzi parameter).
The metric (\ref{metricabella}) for
$\delta_b, a_0 = 0$ is the Schwarzschild metric.

The metric (\ref{metricabella}) has an asymptotic Schwarzschild
core near $ r \approx 0$. To show this property
we develop the metric very close to the point $r\approx 0$ and we consider the
coordinate changing  $R= a_0/ r$.
In the new coordinate the point $r=0$ is mapped
in the point $R=+ \infty$.
The metric in the new coordinates is
\begin{eqnarray}
ds^2= - \bigg(1- \frac{2 m_1}{R} \bigg) dt^2
+ \frac{dR^2}{1- \frac{2 m_2}{R}} + R^2 d \Omega^{(2)},
\label{metricar0b}
\end{eqnarray}
where:  
 $m_1=  A_{\rm Min}/(8 \pi m \gamma^2 \delta_b^2 {\mathcal P}(\delta_b)$,  
$m_2=  A_{\rm Min}  (1 + \gamma^2 \delta_b^2)/(8 \pi m \gamma^2 \delta_b^2 {\mathcal P}(\delta_b))$.

\begin{figure}[tbp]
\leavevmode\hbox{\epsfxsize=12.5 cm
   \epsffile{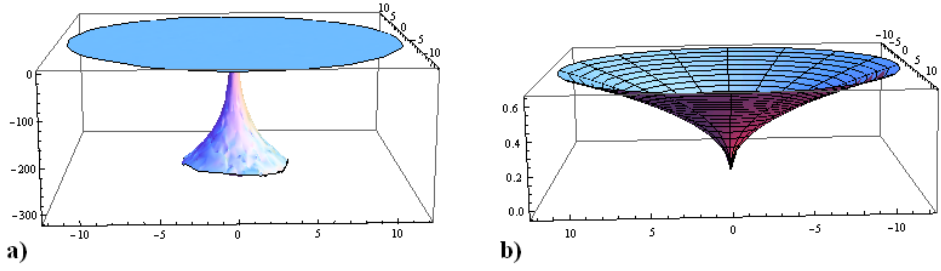}}
\caption{Embedding diagram of a spatial slice just outside the horizon of a 0.005 Planck mass ($\approx 100 ng$) black
hole. In (a) we have the LQBH with metric (\ref{metricabella}); in (b) is the Schwarzschild black hole. In
both cases the foliation is done with respect to the time-like Killing vector and the scales are in Planck units. The
lowermost points in each diagram correspond to the horizon (the outer horizon in the LQG case).}
\label{smalllqgbh}
\end{figure}

For small $\delta_b$ we obtain $m_1\approx m_2$ and (\ref{metricar0b}) converges to
a Schwarzschild metric of mass
$M \approx A_{\rm Min}/ 2 m \pi \gamma^4 \delta_b^4$. We can conclude the space-time
near the point $r\approx 0$ is described by an effective Schwarzschild metric
of mass $M\propto A_{\rm Min}/m$
in the large distance limit $R\gg M$.
An observer in the asymptotic region $r=0$ experiences a Schwarzschild metric
of mass $M\propto a_0/m$.
It is shown in \cite{RNR} that
  a massive particle can not reach
  $r=0$ in a finite proper time.
   The space-time structure of the semiclassical solution is given in Fig.\ref{penrose}.
  \paragraph{Selfduality}
 In this section we explicitly show that the black hole solution obtained in LQG is {\em selfdual}
 in the sense the metric is invariant under the transformation
 $r\rightarrow a_0/r$.
  The self-dual transformation will transform the relevant quantities as: 
 $r \rightarrow R= a_0/r$, 
$r_+ \rightarrow R_- = a_0/r_+$,
$r_- \rightarrow R_+ = a_0/r_-$,
$r_{\star} \rightarrow R_{\star} = a_0/r_{\star}$,
 (note that $R_+ > R_-$ $\forall \delta_b$ because ${\mathcal P}(\delta_b) <1$).
If we apply to this transformation to the metric (\ref{metricabella}), we obtain
\begin{eqnarray}
ds^2 = -\frac{ (R- R_+) (R-R_-)(R+ R_{\star})^2 }{R^4 + a_0^2}dt^2 
%
+\frac{dR^2}{\frac{(R-R_+)(R-R_-)R^4}{(R+ R_{\star})^2 (R^4 + a_0^2)}} 
+\Big(\frac{a_0^2}{R^2} +
R^2\Big) d\Omega^{(2)},
\label{metricabellad}
\end{eqnarray}
where we have complemented the transofmation $r\rightarrow a_0/r$ with a rescaling of the time coordinate
$t \rightarrow {\mathcal P}(\delta_b) ({r_+^{3/2} r_-^{1/2}}/a_0) \, t$.
It is evident from the explicit form (\ref{metricabellad}) that the metric is {\em selfdual}.
 We can define the selfdual 
 radius 
 identifying $R=a_0/r$, $r^{sd}=\sqrt{a_0}$. 
\paragraph{Ultra-light LQBHs}
Outside the exterior horizon, the LQBH 
metric (\ref{metricabella}) differs from the Schwarzschild metric only
by Planck size corrections. As such, the exterior of heavy LQBHs (where by ``heavy" we mean significantly
heavier than the Plank mass) is not qualitatively different than that of a
Schwarzschild black hole of the same mass. 
In order to see a qualitative departure from the Schwarzschild black hole outside the horizon we must consider the
``sub-Planckian" regime. 
Due to Planck scale corrections the radius of the 2-sphere is not $r$, like is the case for
the Schwarzschild metric, but looking at (\ref{metricabella}) we see that the radius of the 2-sphere is
$\rho^2 = r^2 + a_0^2/r^2$.
We see that $\rho$ has a bounce at $r = \sqrt{a_0}$ which comes from LQG having a discrete area spectrum and thus a
minimal area (here $8 \pi a_0 $).  If the bounce happens before the outer horizon is reached, the outer horizon will be
hidden behind the Planck-sized wormhole created where the bounce takes place.  As a consequence of this, even if the
horizon is quite large (which it will be if $m<<m_P$) it will be invisible to observers who are at $r>\sqrt{a_0}$ and
who cannot probe the Planck scale because these observers would need to see the other side of the wormhole which has a
diameter of the order of the Planck length.  From this we conclude that to have this new phenomenon of hidden horizon
we must have $2m= r_+ < \sqrt{a_0}$, or $m<\sqrt{a_0}/2$. We illustrate this phenomenon with the embedding diagrams of
a LQBH of mass $m=4\pi \sqrt{a_0}/100$ in Fig.\ref{smalllqgbh}a
which
can be contrasted with the embedding diagram of the Schwarzschild black hole of the same mass in
Fig.\ref{smalllqgbh}b.

The formation of such ultra-light LQBHs is also of interest. For the Schwarzschild black hole, black hole formation occurs once a critical density is reached, i.e. a mass $m$ is localised inside a sphere of area $4\pi (2m)^2$. The ``heavy" LQBH is analogous: to create it we must achieve a critical density, putting a mass $m\geq \sqrt{a_0}/2$ inside a sphere of area $4\pi [(2m)^2 + a_0^2/(2m)^2]$. The requirement for the formation of an ultra-light LQBH is something else altogether because of the worm-hole behind which it hides: we must localise mass/energy (a particle or a few particles), irrespective of mass as long as the total mass satisfies $m< \sqrt{a_0}/2$) inside a sphere of area $8\pi a_0 $ as this ensures that the mass will be inside the mouth of the wormhole. Because $A_{\rm Min}\geq 5 l_P^2$ for any natural $\beta$ at the currently accepted value of the Immirzi parameter, there does not seem to be any semi-classical impediment to doing that. Hence it should be possible to create ultra-light black holes.

 \paragraph{``Particles-Black Holes" Duality}
In this short section we want to to emphasise the physical meaning of the duality
emerging from the self-dual metric analysed in the paper.
The metric (\ref{metricabella})
describes a space-time with two asymptotic regions, the $r\rightarrow +\infty$ ($\equiv I^+$)
region and the $r\rightarrow 0$ ($\equiv I^0$) region. Two observers in the two regions
see some metric but they perceive two different masses. The observer in $I^+$
perceives a mass $m$, the observer in $I^0$ a mass $M\propto a_0/m$. Physically
any observed supermassive black hole in $I^+$ is perceived as a
an ultra-light ($m\ll m_P$) ``particle" in $I^0$ and vice versa.
The ultra-light ``particle" is confined beyond the throat as discussed in the previous
section because if $m\ll m_P$ then $r_+\ll \sqrt{a_0}$,
which is the throat radius or equivalently the self-dual radius.
This property of the metric leaves open the highly speculative possibility of having a
{\em ``Quantum Particle-Black Hole" Duality}, in fact
a particle with $\lambda_c  \approx \hbar/2m \gg l_P$ could have sufficient space in $r <r_+$
because the physical quantity to compare with $\lambda_c$ is
$D = 2 [(2 G_N m)^2 + a_0^2/(2 G_N m)^2]^{1/2}$ 
and $D \gtrsim \lambda_c \,\, \forall \,\, m$. If $\beta = 4$, $D > \lambda_c$
(it is sufficient that $\beta > 2.43$).
In this way is possible to have a universe dispersed of ultra-light particles ($m\ll m_P$) but
confined inside a sub Planck region and then with a very small cross section.

\section{Phenomenology}
In this section we study the phenomenology of LQBHs and a possible
interpretation of the dark matter and UHECRs in terms ultra-light LQBHs.

\paragraph{Thermodynamics} \label{LQBHT} 

The Bekenstein-Hawking temperature is given in terms of the surface gravity
$\kappa$ by $T= \kappa/2 \pi$. 
Using the semiclassical metric 
we can calculate the surface gravity
in $r = 2m$ 
and then the temperature.
The entropy is defined by $S_{BH}=\int dm/T(m)$ and can be expressed as a function of 
 the event horizon area using the relation: 
 $A=\int d \phi d \theta \sin \theta \, p_c(r)|_{r = 2m}=16 \pi m^2 +  A_{\rm Min}^2/(64 \pi m^2)$.
The results for $T(m)$ and $S(A)$ are:
\begin{eqnarray}
T(m) = \frac{128 \pi \sigma(\delta_b) \sqrt{\Omega(\delta_b)} \, m^3}{1024 \pi^2 m^4 + A_{\rm Min}^2},
\,\,\,\,\,\, 
S= 
 \pm \frac{\sqrt{A^2 - A_{\rm Min}^2}}{4 \sigma(\delta_b) \sqrt{\Omega(\delta_b)}},
\label{Temperatura}
\end{eqnarray}
where $\sigma(\delta_b) \, \Omega(\delta_b) = 16 (1 + \gamma^2 \delta_b^2)^{3/2}/(1 +
     \sqrt{1 + \gamma^2 \delta_b^2})^4$.
The temperature coincides with the Hawking temperature
in the large mass limit and the entropy is positive for $m>\sqrt{a_0}/2$, and negative otherwise.

\begin{figure}[tbp]
\leavevmode\hbox{\epsfxsize=8.5 cm
   \epsffile{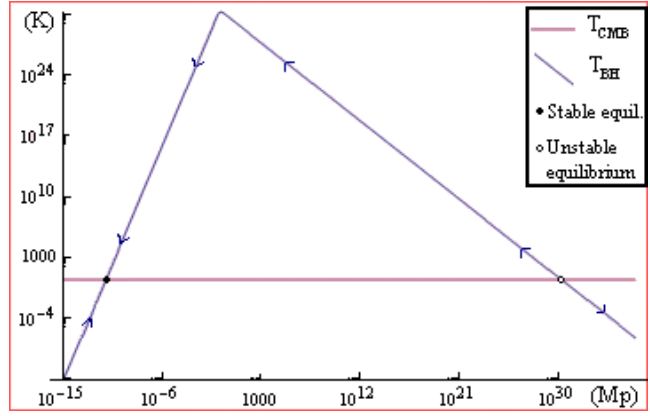}}
\caption{Log-log graph of $T(m)\approx (2m)^3/\{4\pi [(2m)^4+a_0^2]\}$ 
in units of Kelvin and Planck masses. The constant line denotes the temperature of the CMB radiation; above this line the black hole is hotter than the CMB and so it will lose more energy than it gains, below this line the black hole is colder than the CMB and so it will absorb more CMB radiation than it will emit radiation, thereby gaining mass. The arrows on the temperature curve denote in which direction the black hole will evolve through thermal contact with the CMB.  At the two points where the temperature curve intersects the constant $T_{CMB}$ curve, the black hole is in thermal equilibrium. }
\label{logtempera}
\end{figure}

\begin{figure}
 \hspace{-0.8cm}
  \includegraphics[height=5.00cm]{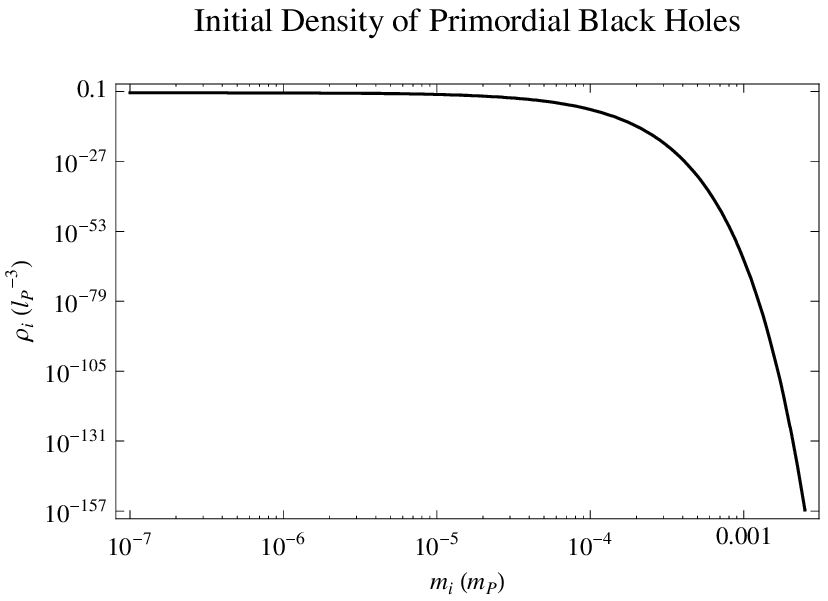}
   \includegraphics[height=4.75cm]{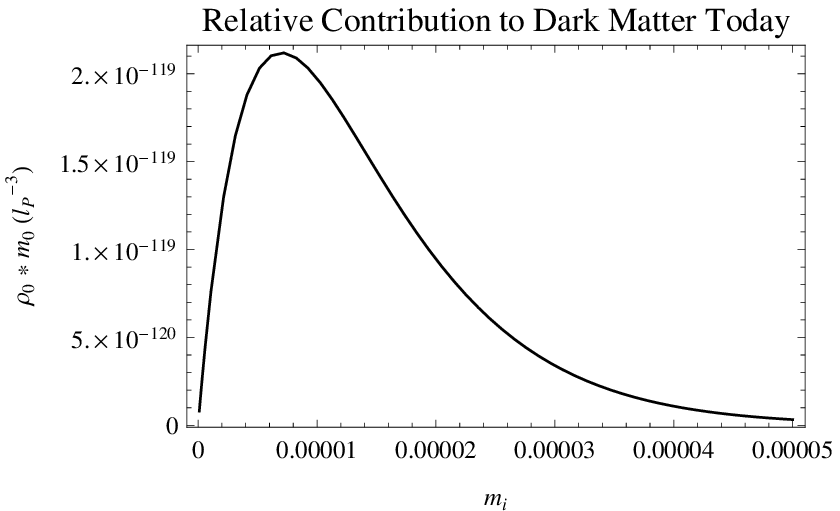}
 \caption{{\em Plot on the left}. The initial density of primordial black holes as given by $\rho_i$ as a function of the initial mass of the black hole. Both the mass $m$ and the temperature are in Planck units. Here we used $\beta=4$ and $T_{eq}= 13\%\times 10^{15}$GeV. The choice of $T_{eq}$ is significant here because the density is very sensitive to $T_{eq}$.
{\em Plot onthe right}. 
This graph shows the current mass density of black holes as a function of their initial mass $m_i$. $\rho_0 (m_0)$ is the current number density of black holes of mass $m_0$, so
$\rho_0=\rho_{i} \, (a_i^3/a_0^3)$. Because, for all practical purposes, $m_0=m_i$, the area under the curve is the present matter density due to LQBHs. If that density is equal to $0.22\rho_{crit}$, the LQBHs will account for all dark matter.  From this graph, we see that at present times, LQBHs mass density is entirely dominated by black holes which had an initial mass of about $ 10^{-5} m_P$. In this graph we have used $\beta=4$ (the graph is not very sensitive to this choice) and $T_{eq}= 13\%\times 10^{15}$GeV (the numerical values of the graph vertical axis are sensitive to this value but location of the peak and the general shape of the graph are not).}
\label{maxrhomod}
  \end{figure}

\paragraph{Ultra-light LQBHs as Dark Matter}
It is interesting to consider how the ultra-light LQBHs might manifest themselves if they do exist in nature. Because they are not charged, have no spin, and are extremely light and have a Planck-sized cross-section (due to their Planck-sized wormhole mouth), they will be very weakly interacting and hard to detect. This is especially true as they need not be hot like a light Schwarzschild black hole, but they can be cold.
 It is thus possible, if they exist, that ultra-light LQBHs are a component of the dark matter. In fact, due to (\ref{Temperatura}), one would expect that all light enough black holes would radiate until their temperature cools to that of the CMB, at which point they would be in thermal equilibrium with the CMB and would be almost impossible to detect. Rewriting (\ref{Temperatura}) for small ${\mathcal P}(\delta_b)$ we get: 
$T(m)\approx (2m)^3/\{4\pi [(2m)^4+a_0^2]\}$.
We thus see emerge a new phenomenon that was not present with Schwarzschild black holes: a black hole in a stable thermal equilibrium with the CMB radiation. In the Schwarzschild scenario, it is of course possible for a black hole to be in equilibrium with the CMB radiation, this happens for a black hole mass of $4.50\times10^{22}$ kg (roughly $60\%$ of the lunar mass). This equilibrium is however not a stable one because for a Schwarzschild black hole the temperature always increases as mass decreases
and vice versa, and so if the black hole is a bit lighter than the equilibrium mass it will be a bit hotter than $T_{CMB}$, the temperature of the CMB radiation, and will emit more energy than it gains thus becoming lighter and lighter. Similarly, if the black hole has mass slightly superior to the equilibrium mass, it will be colder than $T_{CMB}$ and thus absorb more energy than it emits, perpetually gaining mass. For the LQBH however, when $m$ is smaller than the critical mass {\sdmass}
of the order of the Planck mass, the relationship is reversed and the temperature increase monotonically with the mass, this allows for a stable thermal equilibrium in this region as is shown in
Fig.\ref{logtempera}.
The values of the black hole mass in the two equilibrium
positions in the LQG case are thus: 
$m_{{\rm unstable}} = 4.50\times10^{22} \, {\rm kg}$, 
$m_{{\rm stable}} \approx 10^{-19} \, {\rm kg}$; 
where we have used $\gamma = 0.2375329...$ \cite{gamma} 
and assumed $\beta\approx1$. 
\paragraph{LQBHs Production in the Early Universe and Evaporation}
We can estimate the number of ultra-light LQBHs created as well as the extent of their subsequent evaporation. As exposed in \cite{Kapu}, the probability for for fluctuations to create a black hole is $\exp(-\Delta F / T )$, where $\Delta F$ is the change in the Helmholtz free energy and $T$ is the temperature of the universe. The free energy of a LQBH of mass $m$ is
\bea
\hspace{-0.3cm} F_{BH} = m - T_{BH} S_{BH} = m - \half m \, \left[\frac{16 m^4 - a_0^2}{16 m^4 + a_0^2}\right] , \label{Fbh}
\eea
where $T_{BH}$ and $S_{BH}$ are the temperature and entropy of the black hole respectively. The free energy for radiation inside the space where the black hole would form is: 
$F_{R} = - (\pi^2 \kappa /45) \, T^4 V$, 
where $V$ is the volume inside the 2-sphere which will undergo significant change (i.e. significant departure from the original flatness) in the event of a black hole forming 
($\kappa=1$ if only electromagnetic radiation is emitted and $\kappa =36.5$ if all the particles of the Standard Model (including the Higgs) can be radiated). 
 In the case of a black hole of mass $m\geq \sqrt{a_0}/2$, this is naturally the horizon. Since the horizon has an area of  $4\pi [(2m)^2 + a_0^2/(2m)^2]$, we have that the volume of the flat radiation-filled space in which will undergo the transition to a black hole is $V= (4 \pi/3) [(2m)^2 + a_0^2/(2m)^2]^{3/2}$ 
However, as we saw earlier, for example in 
\Ref{smalllqgbh}, if $m\leq \sqrt{a_0}/2$, a throat of a wormhole forms at $r= \sqrt{a_0}$ and a large departure from flat space is observed. Since the mouth of the worm-hole as area $A_{\rm Min}= 8\pi a_0$ we have that the volume of flat space which will be perturbed to create the black hole is $V= (4 \pi/3) (2 a_0)^{3/2}$.
%
Hence, if we define
$\Delta F = F_{BH} - F_{R}$
to be the difference between the black hole free energy and the radiation free energy inside the volume which is to be transformed, we have, in Planck units, that the density of black holes of mass $m$ coming from fluctuations is of the order of
\bea
\rho(m) \approx \frac{1}{\pi^3}\exp(-\Delta F / T ). \label{deltaq}
\eea
One more subtlety however must be considered. 
Formula \rf{deltaq} is only valid if the universe can reach local equilibrium. If the time scale for the expansion of the universe is much shorter than the time scale for collisions between the particles, the universe expands before equilibrium can take place and so \rf{deltaq}, which requires equilibrium, is not valid. It can be shown \cite{mukhanov}, that local equilibrium is reached for temperatures
$T\ll 10^{15} {\rm GeV}-10^{17}{\rm GeV}$. 
 This means that before the universe cooled down to temperatures below $10^{15}$GeV, the universe expanded too quickly to have time to create black holes from fluctuations in the matter density. The fact that the universe must first cool down to below $10^{15}$GeV before a black holes can be created means that black holes of mass $m$ will not be created at temperature $T_{Max}(m)$ 
 for which the maximum amount of black holes are formed
%
and obtained by varying \Ref{deltaq} with respect to $T$, 
 %
 %
 %
 but rather at temperature $T_{cr}(m)=\min\{T_{Max}(m),T_{eq}\}$ where $T_{eq} \lesssim 
 10^{15}$GeV is the temperature below which local equilibrium can be achieved and thus black holes can be created. As can be seen  from equation from $T_{Max}$ 
 this means that for a significant range of black hole masses, from about $10^{-17} \, m_P$ to $10^{8} \, m_P$,  the maximal density will be created when the universe reaches temperature $T_{eq}$. As it turns out, this range will encompass the quasi-totality of black holes responsible for dark matter or any other physical phenomenon considered in this paper.
 The fact that black holes are created only once the universe has cooled down to $T_{eq}$ entails that the initial density of black holes is
$\rho_{i}(m) \approx \frac{1}{\pi^3}\exp(-\Delta F(m) / T_{cr}(m) )$. 
Graphing  
$\rho_{i}(m)$, 
we see in Fig.\ref{maxrhomod} (plot on the left) that only black holes with an initial mass of less then $10^{-3} \, m_P$ are created in any significant numbers.
  
Once the black holes are formed, the only way they can disappear is through evaporation. 
Using that the power of the thermal radiation (in Planck units) emitted by a black body of surface area $A$ and temperature $T$ obeys the Stefan-Bolzmann law:
$P =(\pi^2 \kappa/60) A \, T^4$, 
we find that, all black holes which initially started
with mass $m_i = 0.001 \, m_P$ are de facto stable: the difference
between the initial mass $m_i$ and the mass of the black hole today $m_0$ satisfies in fact if
$(m_i - m_0)/m_i \approx 10^{-14}$ 
(we have taken $\beta =4$ but the result is not sensitive to the exact value of $\beta$) and for smaller initial masses the difference is even smaller. This means that ultra-light black holes are almost 
stable. 
\paragraph{
Number of e-folds Elapsed Since LQBHs Creation  to Account for Dark Matter}
For all black hole initial mass $m_i$, we know 
what the black hole's current mass is. We also know what the initial concentration of each type of black hole was.
In addition, we know that the current matter density for dark matter is approximately $0.22\rho_{crit}$. If we now suppose that currently, all dark matter is actually composed of ultra-light black holes, we have that 
$\int_{0}^\infty (a(t_i)/a(t_0))^3 m_0(m_i) \rho_{i}(m_i) d m_i = 0.22\rho_{crit},$ 
(Fig.\ref{maxrhomod}, plot on the right),
where $a(t_0)$ is the scale factor of the Universe at present ($t_0$), $a(t_i)$ is the scale factor of the universe when the primordial black holes were created ($t_i$) and so $a(t_i)^3 \rho_{i}(m_i)/a(t_0)^3$ is the current number density of black holes of mass $m_0(m_i)$. Since the scale factor does not depend on $m_i$. 
 $N_e  \approx  85$
 and
 $a_0/a_i \approx 10^{37}$,
where we have used $T_{eq}= 1.3\times 10^{14}$GeV and $\beta=4$ though these last two results are very robust under changes of $T_{eq}$ and $\beta$.

Thus, if we want all dark matter to be explained by ultra-light black holes, 
the ultra-light black holes 
have to be created towards the end of the period of inflation which means that inflation should be going on when the universe is at temperature of the order of $10^{14}$GeV$-10^{15}$GeV, 
this is indeed close to the range of temperatures at which inflation is predicted to happen in the simplest models of inflation. 
\paragraph{LQBHs as Sources for Ultra-Hight Energy Cosmic Rays}
Hot ultra-light black holes are very interesting phenomenologically because there is a chance we could detect their presence if they are in sufficient quantities. 
The mass of ultra-light LQBHs today is $m_0 \approx 10^{24} {\rm eV}$, then we can have emission
of  cosmic rays from those object in our galaxy.

In fact, Greisen Zatsepin and Kuzmin proved that cosmic rays which have travelled over $50\, {\rm Mpc}$ will have an energy less than $6\times 10^{19} {\rm eV}$ (called the GZK cutoff) because they will have dissipated their energy by interacting with the cosmic microwave background \cite{GZK}. However, collaborations like HiRes or Auger \cite{HiRes} have observed cosmic rays with energies higher than the GZK cutoff, ultra high energy cosmic rays (UHECR).  The logical conclusion is then that within a
$50 \, {\rm Mpc}$ radius from us, there is a source of UHECR. The problem is that we do not see any possible sources for these UHECR within a $50 \, {\rm Mpc}$ radius. The ultra-light LQBHs which we suggest could be dark matter do however emit UHECR. Could it be that these black holes not only constitute dark matter but are also the source for UHECR? This is not such a new idea, it has already been proposed that dark matter could be the source for the observed UHECR \cite{Xpart}.

\begin{figure}
 \hspace{-0.15cm}
  \includegraphics[height=4.8cm]{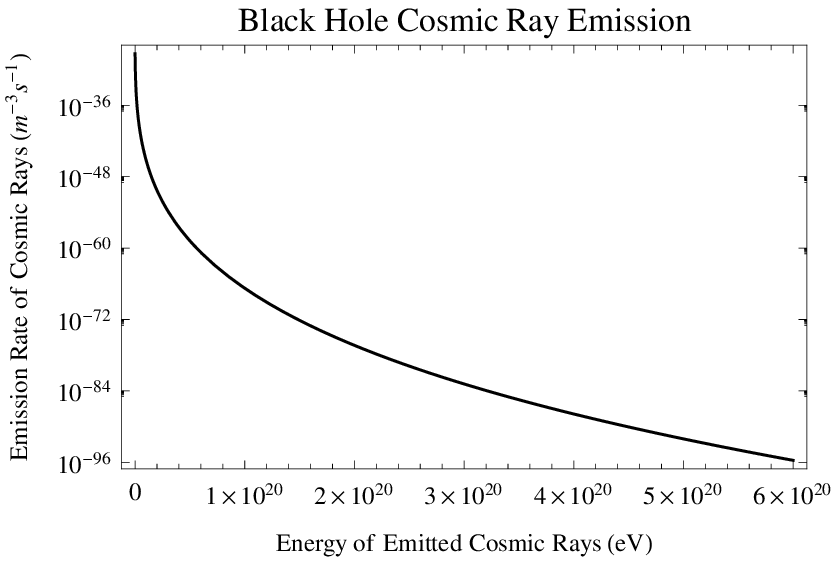}
   \includegraphics[height=4.8cm]{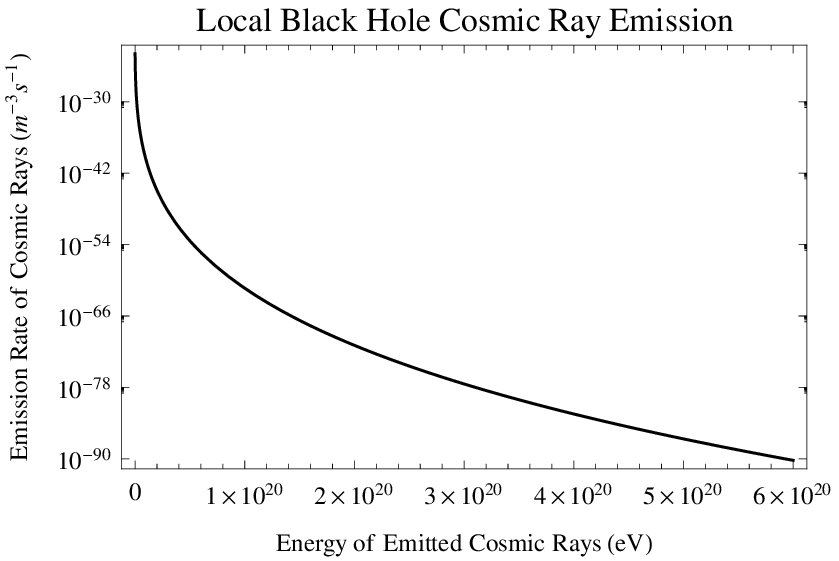}
  \caption{{\em Plot on the left}. The average emission rate of particles by primordial ultra light black holes in the universe given by 
  assuming $\beta=4$ and $T_{eq}=1.3\times10^{14}$GeV.
  {\em Plot on the right}. The local emission rate of particles by primordial ultra light black holes in the Milky Way 
  assuming $\beta=4$ and $T_{eq}=1.3\times10^{14}$GeV.
  }
\label{averageuhecr}
  \end{figure}

In Fig.\ref{averageuhecr} we compare the predicted emissions of UHECR from LQBHs with the observed quantity of UHECR. Auger, for example, observes a rate of UHECR of 
$\sigma_{obs} \approx 10^{-37}  (\text{UHECR particles}) \,\, {\rm s^{-1}} \,\,  {\rm m}^{-3}$. 
If we suppose that that the dark matter in our galaxy is made entirely of LQBHs and that the distribution of LQBHs is the same in our galaxy as the average of the universe only scaled by the appropriate factor to account for the higher dark matter density, we find that this is precisely the observed rate of UHECR one would predict.
This result is very robust for all parameters except for $T_{eq}$ which is very sensitive
and must be fine-tuned to $T_{eq}\approx 13\% \times 10^{15}$GeV. This is in great accordance with 
$T\ll 10^{15} {\rm GeV}-10^{17}{\rm GeV}$.
If $T_{eq}\gg 13\% \times 10^{15}$GeV, then ultra light black holes cannot form the majority of dark matter, because if they did, they would emit much more ultra high energy cosmic rays than we observe. If $T_{eq}\ll 13\% \times 10^{15}$GeV, then it is still possible that ultra light black holes form the majority of dark matter however, they cannot be the source of the ultra high energy cosmic rays which we observe because they will not radiate enough. Only if $T_{eq}\approx 13\% \times 10^{15}$GeV we can have that 
ultra light black holes form majority of dark matter and simultaneously explain UHECRs.

\section*{Conclusions}
In this paper we have studied the new semiclassical 
metric obtained in \cite{RNR}.
The metric has two event horizons that we have defined $r_+$ and $r_-$;
$r_+$ is the Schwarzschild event horizon and $r_-$ is an inside horizon
tuned by the polymeric parameter $\delta_b$.
The solution has many similarities with the Reissner-Nordstr\"om metric
but without curvature singularities anywhere. In particular the region $r=0$ corresponds to
another asymptotically flat region. No massive particle can reach this region in a finite proper time. A careful analysis shows that
the metric has a {\em Schwarzschild core} in $r\approx 0$  of mass $M\propto a_0/m$.
We have studied the black hole thermodynamics : temperature, entropy
and the evaporation process.
The main results are the following.
The temperature $T(m)$ goes to zero
for $m \approx 0$ and reduces to the Bekenstein-Hawking temperature for large
values of the mass
Bekenstein-Hawking.
The evaporation process needs an infinite time. 

From the phenomenological point of view 
ultra-light LQG black holes have the potential to resolve two outstanding problems in physics: what is dark matter, and where do ultra high energy cosmic rays come from. 
If we suppose that the temperature $T_{eq}$ (at the end of inflation) at which local equilibrium of the matter is achieved and thus black holes can be formed from fluctuations of the matter is $13\%$ of $10^{15}$GeV then ultra-light black holes can explain both dark matter and cosmic rays with energies above the GZK cut off.

\paragraph{Acknowledgements}

Research at
Perimeter Institute is supported by the Government of Canada through Industry Canada
and by the Province of Ontario through the Ministry of Research \& Innovation.

\end{document}